\input{aipcheck}

\documentclass[
    ,final            
  ]
  {aipproc}

\layoutstyle{6x9}

\begin{document}

\title{Optically Selected GRB Afterglows}


\author{F. Malacrino}{
  address={ Laboratoire d'Astrophysique Observatoire
  Midi-Pyr\'en\'ees, 14 avenue Edouard Belin, 31400 Toulouse, France}
}

\author{J-L. Atteia\footnote{on behalf of the GRB RTAS collaboration}}{
  address={ Laboratoire d'Astrophysique Observatoire
  Midi-Pyr\'en\'ees, 14 avenue Edouard Belin, 31400 Toulouse, France}
}

\begin{abstract}
Since November 2004, we attempt to detect GRB optical afterglows in near real-time on images
taken at the Canada France Hawaii Telescope within the Very Wide Survey, component of the CFHT
Legacy survey. To do so, a Real Time Analysis System automatically and
quickly analyzes MegaCAM images and extracts from them a list of
photometrically and astrometrically variable objects which is then
validated by a member of the collaboration. Each month, we repeatedly
observe 15 to 30 square degrees down to magnitude i' = 22.5.  A few
objects are classified as candidates and analyzed more deeply, and
statistics are done showing the treatment's performance.
\end{abstract}

\maketitle


\section{The Survey}
The Canada France Hawaii Telescope is a 3.6 m telescope located on the Mauna Kea in Hawaii.
Built in the late 70's, it has been recently equipped with a new
instrument, Megacam, a full square degree CCD imager.

The Very Wide Survey covers 1200 square degrees down to i' = 22.5,
through 3 filters (r', g', i'). Initially conceived to discover and
follow 2000 Kuiper Belt Objects, its observing strategy (Fig 1) is well suited to
detect optical afterglows.

\begin{figure}[!h]
  \includegraphics[height=.15\textheight]{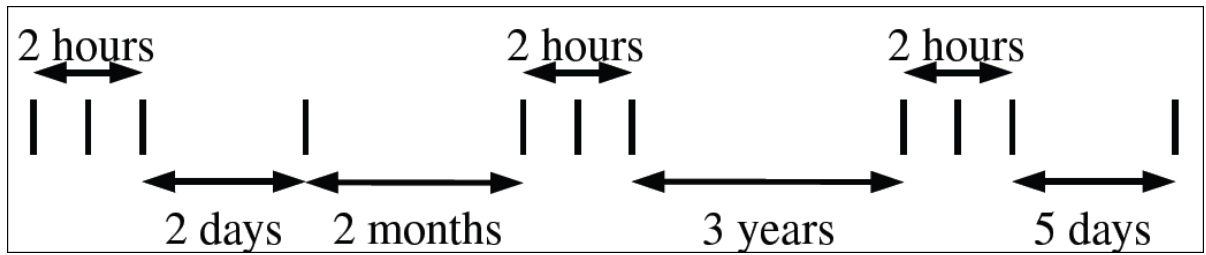}
  \caption{Observational strategy for one field (each vertical line is for one exposure, exposure time depends on the filter, but is typically of the order of one hundred seconds)}
\end{figure}

Each field is observed several times. This recurrence can be used to
compare images between them in order to detect variable or transient
objects.

\section{The Real Time Analysis}
\subsection{Catalog Process}
This part of the process consists of reducing the useful information
from 700 Mo to a few ten Mo and to prepare all the useful data for the
comparison process. The pipeline automatically checks the presence of
a new image as soon as it has been pre-processed by Elixir and starts
the treatment for each of the 36 CCDs.

It consists of converting FITS images to JPEG format,
extracting FITS headers and creating a catalog of objects, containing
their main characteristics. These objects are then sorted according to
their astrophysical properties (stars, galaxies, cosmic rays, etc..)
and astrometically matched with the USNO catalog. Results are finally
summarized on a web page to be checked and validated by a
collaboration member. Images treated in this way can then be used in
the comparison process.

\subsection{Comparison Process}
The goal of this process is first to list all the possible comparisons
between images of the same field. To be possible, a comparison must
involve images with the same filter and exposure time. In a second step,
the process checks if the field has already been observed during this
run, and in this case starts the comparison between the two best
quality images of each night.

The comparison process classifies objects into two categories: matched
and single objects. Matched objects are used for photometric
inter-calibration. Asteroids are detected among single objects. At last, a
search of matched objects having significant variation in luminosity
is done.

All the information about variable objects and asteroids is gathered
on a interactive web page. Then a member of the collaboration has to reject false
detections or to validate objects as really variable. This validation
process typically takes 2-3 hours for 1 run.

\begin{figure}[!h]
\centering
  \includegraphics[height=.29\textheight]{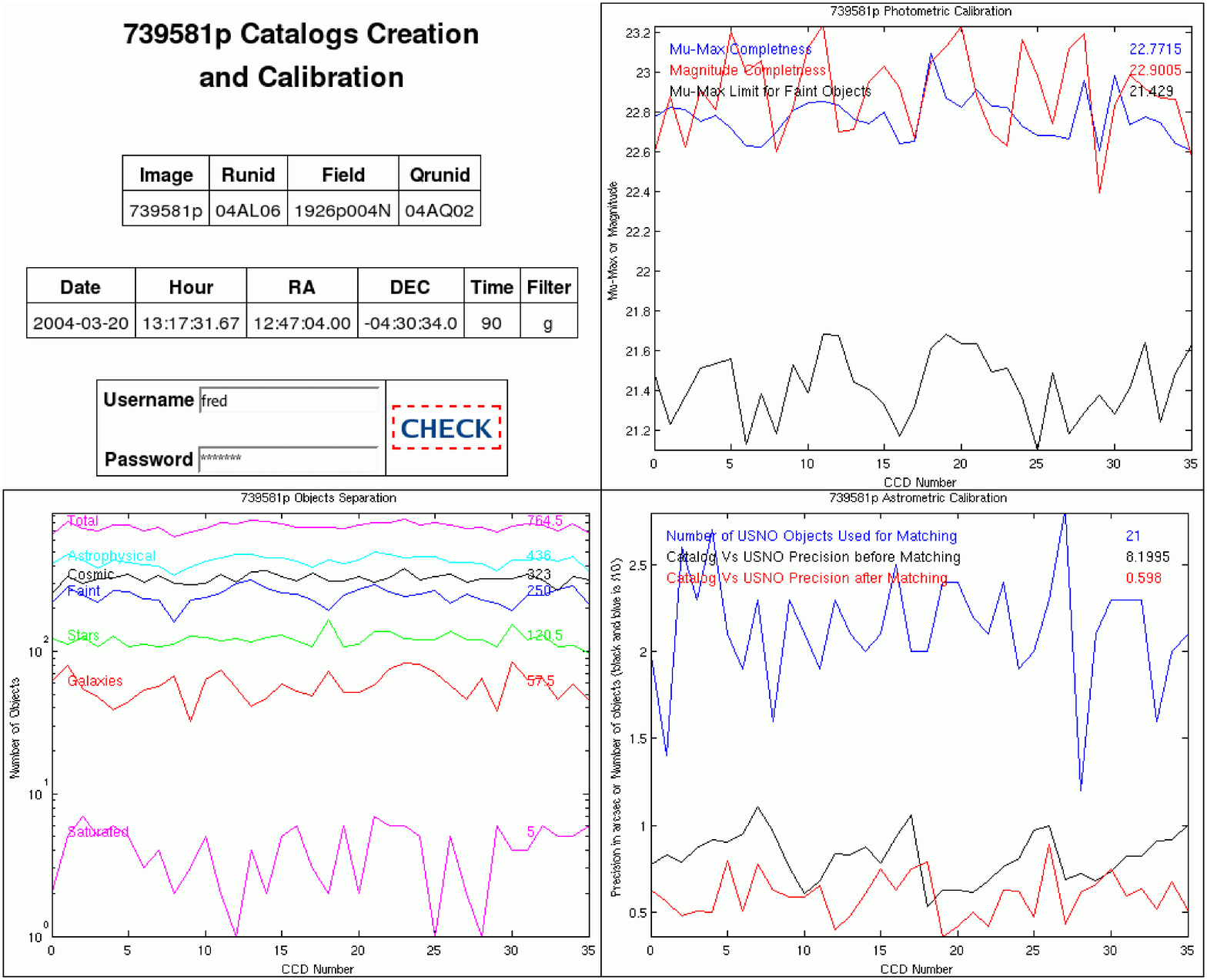}\hfill
  \includegraphics[height=.29\textheight]{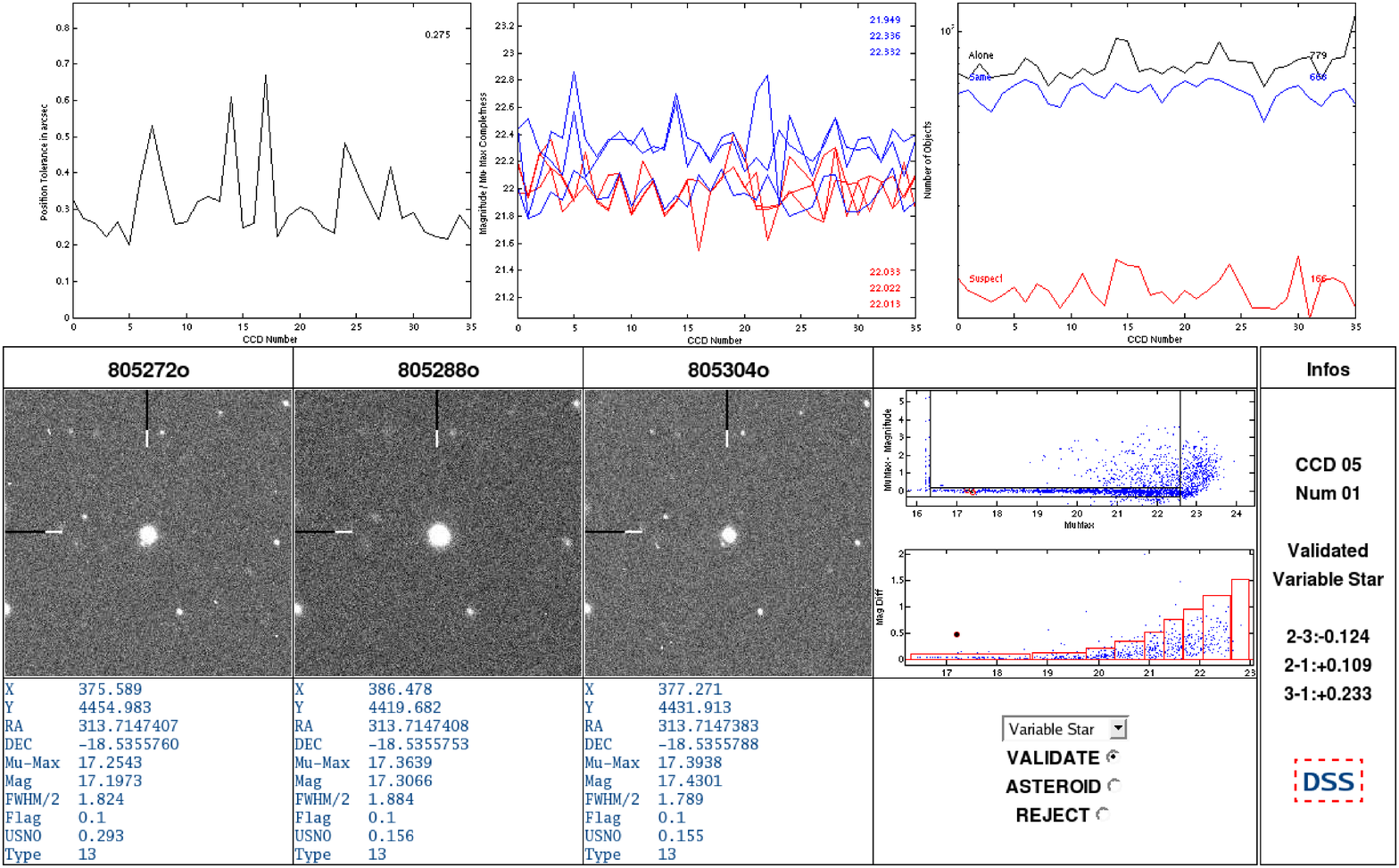}
  \caption{Example of a "catalog creation" web page (left) and a "triple comparison" web page (right)}
\end{figure}

\section{The Post Analysis}
At the end of each run, statistics are performed on catalogs,
comparisons and validation. These statistics allow a more
comprehensive view of the treatment and the possibility of comparing
different runs. Ultimately the performances of the survey
will be evaluated on a population of simulated GRB afterglows
to constrain the rate of GRB afterglows.

\paragraph{Catalogs}
Table 1 shows statistics for the catalog process of the
MegaCAM run 05BQ07. The Very Wide survey usually works by sets of 16
or more fields, that's why images are classified by sets. We can
notice that i' filter images contain twice more objects than r'
filter ones, although the completness magnitudes are almost the
same. It can also be noted that the last i' filter set is worse
compared with the first two due to undefringe images.

\begin{table}[!h]
\begin{tabular}{cccccccc}
\hline
    \tablehead{1}{c}{b}{Filter}
  & \tablehead{1}{c}{b}{Set}
  & \tablehead{1}{c}{b}{Date}
  & \tablehead{1}{c}{b}{Number of\\Images}
  & \tablehead{1}{c}{b}{Square\\Degrees}
  & \tablehead{1}{c}{b}{USNO\\Precision}
  & \tablehead{1}{c}{b}{Completness\\Magnitude}
  & \tablehead{1}{c}{b}{Objects per\\Square Degrees}\\
\hline
r' & 1 & 2005-10-25 & 18 & 16.018 & 0.60 & 22.53 & 15652\\
r' & 2 & 2005-10-25 & 16 & 14.288 & 0.59 & 22.40 & 14415\\
r' & 3 & 2005-10-25 & 17 & 15.165 & 0.60 & 22.34 & 14139\\
r' & 4 & 2005-10-26 & 17 & 15.266 & 0.60 & 22.19 & 12613\\
i' & 1 & 2005-10-26 & 16 & 14.438 & 0.62 & 22.32 & 28926\\
i' & 2 & 2005-10-26 & 16 & 14.438 & 0.62 & 22.39 & 30221\\
i' & 3 & 2005-10-26 & 17 & 15.341 & 0.62 & 22.10 & 28499\\
\hline
\end{tabular}
\caption{Statistics for 05BQ07 catalogs}
\end{table}

\paragraph{Comparisons}
Using these catalogs, 32 triple comparisons have been processed, 16
for each filter. Triple comparisons involves images of the same night.
r' filter triple comparisons are very impressive, with only 0.05 \%
of photometrically variable objects and many asteroids detected down to
magnitude r' = 22.1. Unfortunately, the bad third image of the i' filter set leads to a
huge number of photometrically variable objects, and this set of
triple comparisons hasn't been validated by the team due to many false
detections.
\begin{figure}[!h]
  \includegraphics[height=.22\textheight]{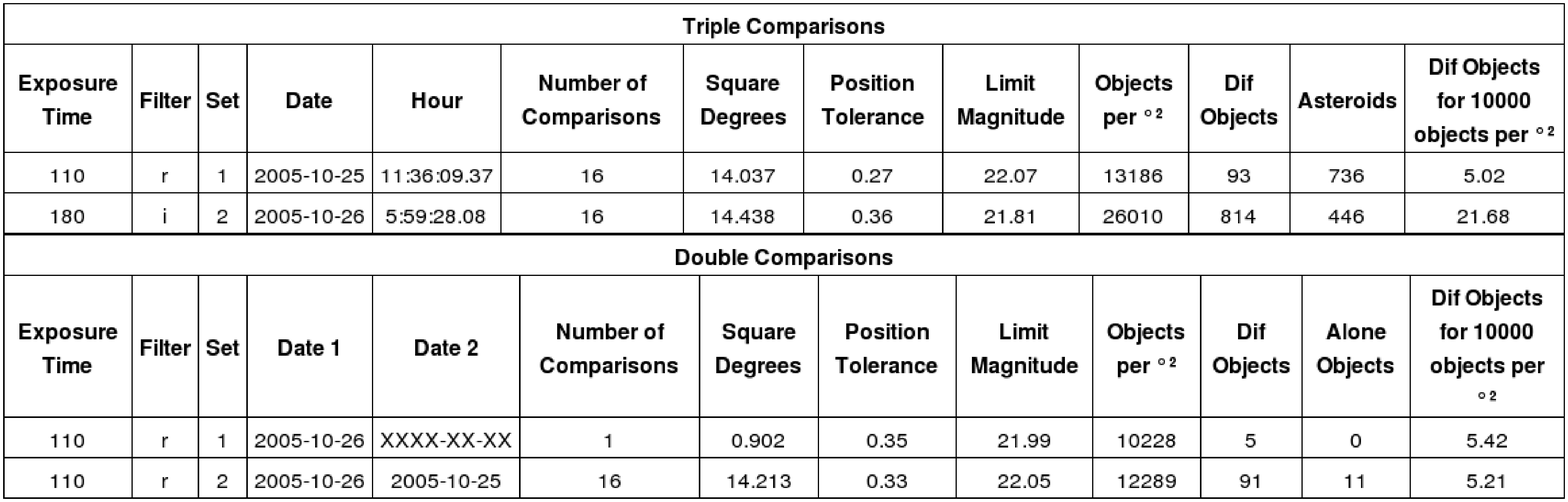}
  \caption{Statistics for 05BQ07 comparisons}
\end{figure}

Double comparison statistics only concern r' filter images. The
first set is insignificant, as it contains only one double comparison
of 2 images of the same night. On the contrary, the second set looks
like what we get in case of perfect double comparisons. Once again,
the process selects approximatively 0.05\% of luminosity variable
objects down to magnitude r' = 22, and 11 objects which were classified
as matched in the corresponding triple
comparison, and are not present in the image taken one day later. The behaviour of these objects
looks like the expected behaviour of GRB afterglows, showing that we can indeed detect GRB afterglows.

\paragraph{Validation}
In the last part, we combine triple and double comparisons, and their corresponding validations, in
order to compute statistics for asteroids and variable objects per field, as we
consider that fields observed 2 days apart are not independent.

\begin{figure}[!h]
  \includegraphics[height=.07\textheight]{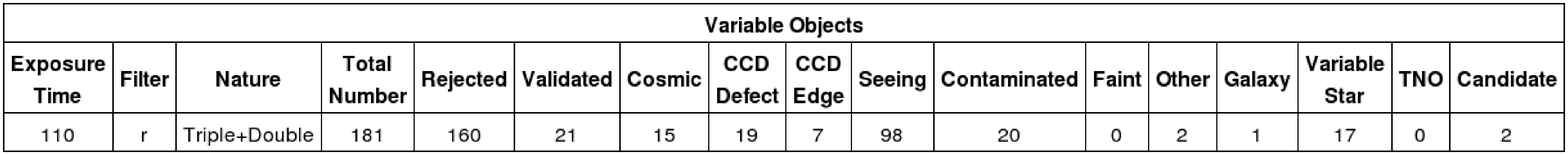}
  \caption{Statistics for 05BQ07 variable objects validation}
\end{figure}

Only 12\% of objects classified as variable by the process
are really variable, and most of them are variable stars. More than 50\%
of the detections are due to seeing variations, despite corrections
applied by the process. 5 interesting candidates have been identified
which were all discarded by comparison with previous images of the
field.




\section{Conclusion}
We have shown that optical afterglow detection is possible using
Megacam images from the Very Wide survey at the CFHT and a real time
comparison pipeline. About 1 to 10 afterglows per year can be
reasonably expected. The real time process has started one year ago,
and 12 runs have been succesfully processed, but unfortunately without
any strong afterglow candidate. Now, we have started the complete
reprocessing of the whole Very Wide survey runs since the beginning of
the legacy survey in April 2003. This study will allow us to compute a huge database
of objects which will be very useful to plot luminosity function of
the most interesting objects, and to strongly constrain the rate of
orphan GRB afterglows and their beaming factor. Our work is continuing with
new images taken every month, and we recently submitted a proposal for the next
semester, which is specifically dedicated to the search of orphan GRB
afterglows and will have a efficiency 2-3 times greater than the Very
Wide Survey for this task.

Please visit \url{http://www.cfht.hawaii.edu/~grb} for more informations on this research.


\begin{theacknowledgments}
The authors gratefully acknowledge the support of the CFHT staff
for the steady operation of the RTAS.
\end{theacknowledgments}






\end{document}